\documentclass[prd,a4paper,showpacs,amsmath]{revtex4}
\usepackage{bm}
\usepackage{color,xcolor}
\usepackage{amsmath}
\usepackage{amssymb}
\usepackage{amsfonts}
\usepackage{keyval,graphicx}
\usepackage{slashed}
\usepackage{epstopdf}
\usepackage{isotope}
\usepackage{dcolumn}

\begin{document}

\title{Photoproduction of hidden charm pentaquark states  $P_c^+(4380)$ and $P_c^+(4450)$ }

\author{Qian Wang$^1$\footnote{q.wang@fz-juelich.de}, Xiao-Hai Liu$^2$\footnote{liuxh@th.phys.titech.ac.jp},
and Qiang Zhao$^3$\footnote{zhaoq@ihep.ac.cn}}

\affiliation{$^1$   Institut f\"{u}r Kernphysik, Institute for Advanced Simulation, and J\"ulich Center for Hadron
Physics, D-52425 J\"{u}lich, Germany }

\affiliation{$^2$ Department of Physics, H-27, Tokyo Institute of Technology, Meguro, Tokyo 152-8551, Japan}

\affiliation{$^3$ Institute of High Energy Physics and Theoretical Physics Center for Science Facilities,
        Chinese Academy of Sciences, Beijing 100049, China}

\begin{abstract}
We propose to study the pentaquark candidates of $P_c^+(4380)$ and $P_c^+(4450)$ in $J/\psi$
photoproduction and look for further experimental evidence for their nature. Since the photoproduction process does not satisfy
the so-called ``anomalous triangle singularity" condition their presence in $J/\psi$ photoproduction would conclude that they should
be genuine states and provide further evidence for their existence.
\end{abstract}

\date{\today}
\maketitle

\section{Introduction}

Since the establishment of QCD the study of hadron states beyond the simple quark model, namely the QCD exotic states, has been a topical subject in hadron physics. However, for a long time decisive evidence for QCD exotics has been absent from experimental measurements. Until recently, more and more evidence in the heavy quark sector indicates possible candidates for the QCD exotics which are named as ``XYZ" states. Very recently, the LHCb Collaboration reported observation of two multiquark states, $P_c^+(4380)$ and $P_c^+(4450)$, in the invariant mass spectrum of $J/\psi p$ in the decay of $\Lambda_b^0\to J/\psi K^- p$~\cite{Aaij:2015tga}. The masses of these two states are $4380\pm 8 \pm 29$ MeV and $4449.8\pm 1.7\pm 2.5$ MeV, respectively, and their widths are $205\pm 18\pm 86$ MeV and $39\pm 5\pm 19$ MeV, respectively. Their quantum numbers are determined by the partial wave analysis and the spins are $3/2$ and $5/2$ with opposite parities. Since their decays into $J/\psi p$ contain at least five quarks, they are ideal candidates for pentaquark states of $uudc\bar c$. This experimental result immediately initiates a lot of discussions and studies. Although there are immediate publications claiming their dynamic properties as a multiquark state~\cite{Chen:2015loa,Chen:2015moa,Karliner:2015ina,Roca:2015dva,Feijoo:2015cca,Maiani:2015vwa,Meissner:2015mza,Anisovich:2015cia,Lebed:2015tna}, it was also investigated in Refs.~\cite{Guo:2015umn,Liu:2015fea,Mikhasenko:2015vca} that the anomalous triangle singularity (ATS) may produce threshold enhancements to mimic the behavior of genuine states or to contribute on top of the genuine states.

It is interesting to note that since these two states were observed in their decays into $J/\psi p$, it is natural to expect that they can be produced in $J/\psi$ photoproduction via the $s$ and $u$-channel process. This motivates us to investigate these two pentaquark candidates in $J/\psi$ photoproduction. In particular, if they are genuine states, they should be created in $J/\psi$ photoproduction, while if they are the ATS enhancement, they will not appear in $\gamma p\to J/\psi p$ since the ATS condition cannot be satisfied here. Therefore, their production in $J/\psi$ photoproduction can provide direct evidence for their nature as genuine states. This process can be studied at JLab in the near future. Another interesting process is $e^+ e^-\to J/\psi p\bar p$ where the production of the pentaquark states in the invariant spectrum of $J/\psi p$ is also free of the ATS ambiguity. Such a possibility can be investigated by Belle or future Belle-II.

There have been studies of exotic hadrons via meson photoproduction  \cite{Liu:2008qx,Brodsky:2015wza}. For instance, it was first proposed in Ref.~\cite{Liu:2008qx} to study the tetraquark candidate $Z(4430)$ in meson photoproduction due to its strong coupling to $\psi'$. This is an alternative way to probe the property of $Z(4430)$. Similar to $Z(4430)$, the pentaquark candidates $P_c^+(4380)$ and $P_c^+(4450)$ observed in their decays into $J/\psi p$ as genuine states should have sizeable couplings to $\gamma p$. One can expect that their decays into $\gamma p$ should be through the $c\bar c$ annihilations, or it means that such hidden-charm baryons can be created in $J/\psi$ photoproduction via the $s$-channel excitations. Recent studies of hidden-charm pentaquark states can be found in the literature~\cite{Wu:2012xg,Huang:2013mua} where, however, only low spin states were investigated. The LHCb analysis  favors that  $P_c^+(4380)$ and $P_c^+(4450)$ to have higher spins of either $3/2$ or $5/2$. Their production mechanism in $\gamma p$ scatterings should be investigated.

The available data for $J/\psi$ photoproduction show that the cross section is dominated by the diffractive process, i.e. the Pomeron exchanges. This is a $t$-channel process featured by the forward peaking in the differential cross section. There have been tremendous studies of the diffractive process in the literature (see e.g. Ref.~\cite{Donnachie:2002en}). Due to the dominance of the diffractive process, it gives us an opportunity for searching for the hidden-charm pentaquark states in $J/\psi$ photoproduction which will be produced strongly via the $s$-channel transition and should dominate at off-forward kinematics.

In this work we study the photoproduction of the  $P_c^+(4380)$ and $P_c^+(4450)$ off the proton, since this process cannot satisfy the ATS condition which makes it is a good place to distinguish a genuine state from the ATS kinematic effect. We include the $t$ channel diffractive Pomeron exchanges and the $s$-channel pentaquark productions. By assuming that the pentaquark states dominantly decay into $J/\psi p$, we can extract their couplings to $J/\psi p$ by different possible quantum numbers. We then provide the different cross sections where the angular distributions for different quantum numbers will allow us to learn their nature from experiment. As follows, we will present the formalism of our model in Section II. The numerical results follows in Section III and a brief summary will be given in Section IV.

\section{Formalism}
In this section, we present detailed formulas for the dominant diffractive Pomeron exchange in the $t$ channel and the pentaquark excitation in the $s$ channel in $J/\psi$ photoproduction.

\subsection{Pomeron exchange}

\begin{figure}
\centering
\includegraphics[width=0.6\textwidth]{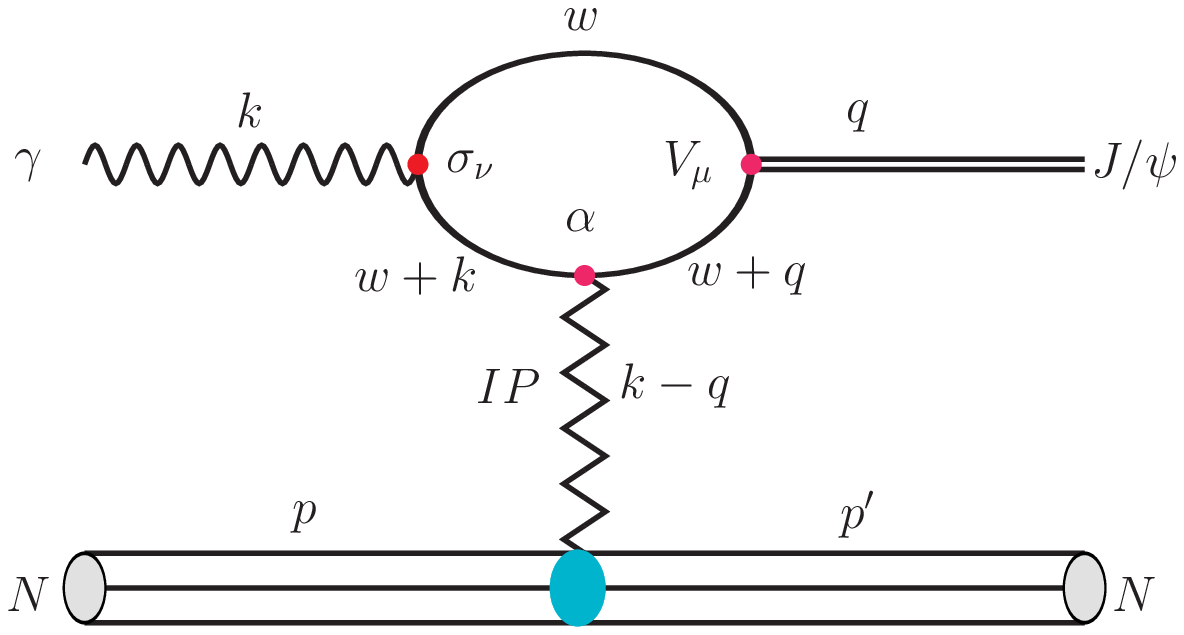}
\caption{The diffractive process via Pomeron exchange in  $J/\psi$
photoptroduction.  }
\label{fig-2}
\end{figure}

In $\gamma p\to J/\psi p$, the dominant contribution is the diffractive process which is accounted for by the Pomeron exchange model~\cite{donnachie,pichowsky,laget}. In this scenario the Pomeron mediates (Fig.~\ref{fig-2}) the long range interaction between a confined quark and a nucleon. Although the nature of Pomeron (generally interpreted as multi-soft-gluon exchanges) is still unclear, it is a useful tool for describing the diffractive production of neutral vector mesons in the high energy region.

With the same behavior of a
$C=+1$ isoscalar photon, the coupling between Pomeron and quark and/or
antiquark contributes a Lorentz index $\gamma_{\alpha}$. So the
Pomeron-nucleon coupling is expressed as
\begin{equation}
{\cal F}_{\alpha}=3\beta_{0}\gamma_{\alpha}f(t).
\end{equation}
Where $\beta_{0}$ gives the strength of the coupling of a single
Pomeron to a light constituent quark, and $f(t)$ represents the
structure function of the nucleon which is expressed as follows
\begin{equation}
f(t)=F_{1}(t)=\frac{(4M_{P}^{2}-2.8t)}{(4M_{P}^{2}-t)(1-t/0.7)^2}  ,
\end{equation}
with $t$ the $t$-channel momentum transfer squared.

Taking the on-shell approximation for the quark loop, a gauge-fixing scheme~\cite{Titov:1998bw} is adopted and the $J/\psi \gamma {\cal P}$ vertex has the following structure:
\begin{equation}
T^{\alpha,\mu\nu} = ( k + q )^\alpha g^{\mu\nu} - 2k^\mu
g^{\alpha\nu}.
\end{equation}

The amplitude of Pomeron exchange is
\begin{equation}
\mathcal{M}_{1}=2\beta_{0}\tilde{T}^{\alpha,\mu\nu}\epsilon_{J/\psi\mu}\epsilon_{\gamma\nu}\bar{u}(p^{\prime})\mathcal{F}_{\alpha}(t)u(p)\zeta_{p}(s,t),
\end{equation}
where $\zeta_{p}(s,t)$ is Regge trajectory of the Pomeron and it is written as follows
\begin{equation}
\zeta_{p}(s,t)=-i(\alpha^{\prime}s)^{\alpha(t)-1} \ ,
\end{equation}
where $\alpha(t)=1+\epsilon^{\prime}+\alpha^{\prime} t$. For any off-shell quark line a form factor, $\frac{\mu_{0}^{2}}{\mu_{0}^{2}+m^2-p^{2}}$, is
introduced, where $p$ is the four momentum transfer in the quark line and $m$ is the mass of the quark. Equally it will provide a form factor
\begin{eqnarray}
\tilde{\cal F}(t)\equiv \frac{2 \beta_c 4 \mu_0^2}{(m_{J/\psi}^2-t)(2\mu_0^2+m_{J/\psi}^2-t)}
\end{eqnarray}
for the $\gamma J/\psi p$ vertex. Then, $\tilde{T}^{\alpha,\mu\nu}\equiv T^{\alpha,\mu\nu}\tilde{\cal F}(t)$ is defined. All the parameters used for Pomeron are the same as those in Ref.~\cite{Liu:2008qx}.

\subsection{$s$-channel pentaquark production}

\begin{figure}
\centering
\includegraphics[width=0.9\textwidth]{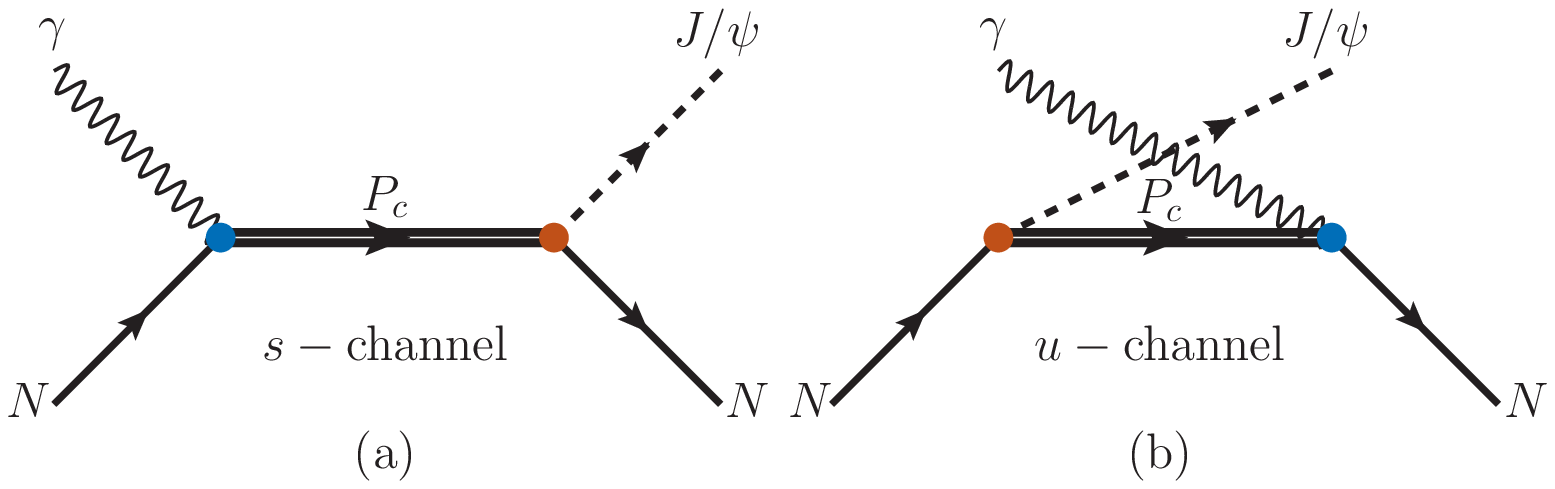}
\caption{The $s$ and $u$-channel production of pentaquark states $P_c^+$ in $J/\psi$
photoptroduction.  }
\label{fig-3}
\end{figure}

The excitation of pentaquark candidates, $P_c^+(4380)$ and $P_c^+(4450)$ in $J/\psi$ photoproduction can occur via both $s$ and  $u$ channel as illustrated by Fig.~\ref{fig-3} (a) and (b), respectively. However, we should note that the $u$-channel contribution is negligible since the intermediate $P_c^+$ will be highly off-shell. 

Since the LHCb analysis suggests that  $P_c^+(4380)$ and $P_c^+(4450)$ favor spin $3/2$ and $5/2$, respectively, but with opposite signs, we then investigate the spin parity assignments of $(3/2^-, 5/2^+)$ and $(3/2^+, 5/2^-)$ for these two states in this study. We adopt the effective Lagrangians from Ref.~\cite{Kim:2011rm} for the $\gamma N P_c^+ $ couplings with $J^P=3/2^\pm$ or $5/2^\pm$ for the $P_c^+$:
\begin{eqnarray}\label{photocoupling}
{\cal L}_{\gamma N P_c}^{3/2^\pm} &=& \frac{i e h_1}{2M_N} \bar N \Gamma_\nu^{(\pm)} F^{\mu\nu} P_{c\mu} -\frac{e h_2}{(2M_N)^2} \partial_\nu \bar N\Gamma^{(\pm)}  F^{\mu\nu} P_{c\mu} + H.c. \ , \\
{\cal L}_{\gamma N P_c}^{5/2^\pm} &=& \frac{e h_1}{(2M_N)^2} \bar N \Gamma_\nu^{(\mp)}\partial^\alpha F^{\mu\nu} P_{c\mu\alpha} -\frac{ie h_2}{(2M_N)^3}\partial_\nu \bar N  \Gamma_\nu^{(\mp)} \partial^\alpha F^{\mu\nu} P_{c\mu\alpha} + H.c. \ ,
\end{eqnarray}
where $P_c$ is the pentaquark fields with spin-parity $J^P=3/2^\pm$ or $5/2^\pm$; $N$ is the nucleon field. The $\Gamma$ matrix  is defined as follows:
\begin{equation}\label{Gamma-matrix}
\Gamma_\mu^{(\pm)}\equiv \left(\begin{array}{c}
\gamma_\mu \gamma_5 \\
\gamma_\mu
\end{array}
\right) \ , \ \ \ \Gamma^{(\pm)}\equiv \left(\begin{array}{c}
\gamma_5 \\
\mathbf{1}
\end{array}
\right) \ ,
\end{equation}
and superscript $\mp$ in the $\Gamma$ matrices means interchange the rows in Eq.~(\ref{Gamma-matrix}). The coupling constants in Eq.~(\ref{photocoupling}) will be related to the couplings for $P_c N J/\psi$ interaction via vector meson dominance (VMD).

We also adopt the effective Lagrangians for $P_c N J/\psi$ couplings from Ref.~\cite{Kim:2011rm} as follows:
\begin{eqnarray}
{\cal L}_{P_c N \psi}^{3/2^\pm} & =&
-\frac{i g_1}{2M_N} \bar N \Gamma_\nu^{(\pm)} \psi^{\mu\nu} P_{c\mu}
-\frac{g_2}{(2M_N)^2} \partial_\nu \bar N\Gamma^{(\pm)} \psi^{\mu\nu} P_{c\mu}
+\frac{g_3}{(2M_N)^2} \bar N \Gamma^{(\pm)}\partial_\nu \psi^{\mu\nu} P_{c\mu} + H.c. \ ,\\
{\cal L}_{P_c N \psi}^{5/2^\pm} & =&
\frac{g_1}{(2M_N)^2} \bar N \Gamma_\nu^{(\mp)} \partial^\alpha \psi^{\mu\nu} P_{c\mu\alpha}
-\frac{i g_2}{(2M_N)^3} \partial_\nu \bar N\Gamma^{(\mp)} \partial^\alpha \psi^{\mu\nu} P_{c\mu\alpha}
+\frac{ig_3}{(2M_N)^3} \bar N \Gamma^{(\mp)}\partial^\alpha\partial_\nu \psi^{\mu\nu} P_{c\mu\alpha} + H.c. \ .
\label{strong-coupling}
\end{eqnarray}

One can see that for these higher partial wave states they can have different coupling structures which will make it difficult to determine their values. However, notice that for $P_c^+(4380)$ and $P_c^+(4450)$ decays into $J/\psi p$, the momentum of the final states are fairly small compared with the nucleon mass. Thus, the higher partial wave terms proportional to $(p/M_N)^2$ and $(p/M_N)^3$ in Eqs.~(\ref{photocoupling}) to (\ref{strong-coupling}) can be neglected. Then, the transition matrix elements to the leading order can be written as
\begin{eqnarray}
{\cal M}^{3/2^\pm} &=& \frac{1}{s-M_{P_c}^2}\frac{eh_1 g_1}{(2M_N)2}\epsilon^*_{\psi\nu} \bar u_N\Gamma_\sigma^{(\pm)} \Delta_{\beta\alpha}(P_c, k+p) \Gamma_\delta^{(\pm)}
(k^\alpha g^{\mu\delta} - k^\delta g^{\alpha\mu}) u_N \epsilon_{\gamma\mu} \ , \\
{\cal M}^{5/2^\pm} &=& \frac{1}{s-M_{P_c}^2}\frac{eh_1 g_1}{(2M_N)4}\epsilon^*_{\psi\nu} \bar u_N q^\sigma(q^\rho g^{\nu\delta}- q^\delta g^{\nu\rho})\Delta_{\rho\sigma; \ \alpha\beta}(P_c, k+p) \Gamma_\lambda^{(\mp)} k^\beta( k^\alpha g^{\mu\lambda}-k^\lambda g^{\alpha\mu}) u_N \epsilon_{\gamma\mu} \ ,
\end{eqnarray}
where $\epsilon_{\gamma\mu}$ and $\epsilon^*_{\psi\nu}$ are the polarization vectors for the initial state photon and final state $J/\psi$, respectively; $\Delta_{\beta\alpha}(P_c, k+p)$ and $\Delta_{\rho\sigma; \ \alpha\beta}(P_c, k+p)$ are the spin $3/2$ and $5/2$ Rarita-Schwinger spin projections for the corresponding baryon states, respectively. They have the following expressions~\cite{Kim:2011rm}:
\begin{eqnarray}
\Delta_{\beta\alpha}(B, p) &= & (\slashed{p} + M_B) \left[-g_{\beta\alpha} +\frac 13\gamma_\beta\gamma_\alpha +\frac{1}{3M_B}(\gamma_\beta p_\alpha-\gamma_\alpha p_\beta) +\frac{2}{3M_B^2} p_\beta p_\alpha\right] \ , \\
\Delta_{\rho\sigma; \ \alpha\beta}(B, p) & =& (\slashed{p} + M_B) \left[\frac 12 (\bar g_{\rho\alpha} \bar g_{\sigma\beta} + \bar g_{\rho\beta} \bar g_{\sigma\alpha}) -\frac 15 \bar g_{\rho\sigma} \bar g_{\alpha\beta} -\frac{1}{10} (\bar\gamma_\rho \bar\gamma_\alpha \bar g_{\sigma\beta} + \bar\gamma_\rho \bar\gamma_\beta \bar g_{\sigma\alpha} + \bar\gamma_\sigma \bar\gamma_\alpha \bar g_{\rho\beta} + \bar\gamma_\sigma \bar\gamma_\beta \bar g_{\rho\alpha}) \right] \ ,
\end{eqnarray}
where $M_B$ and $p$ denote the mass and momentum of baryon $B$ and $\bar g_{\alpha\beta}$ and $\bar\gamma_\alpha$ are defined as the following:
\begin{eqnarray}
\bar g_{\alpha\beta} &=& g_{\alpha\beta}-\frac{p_\alpha p_\beta}{M_B^2} \ ,\\
\bar\gamma_\alpha &=& \gamma_\alpha-\frac{p_\alpha}{M_B^2}\slashed{p} \ .
\end{eqnarray}

The decay of the $P_c^+$ states into $\gamma p$ should be via the $c\bar c$ annihilations. Since these states couple to $J/\psi p$, we assume that
their electromagnetic (EM) couplings are via the $J/\psi$ pole.
We adopt the VMD leading order coupling between a vector meson and photon:
\begin{equation}
{\cal L}_{V\gamma} = \sum_V\frac{eM_V^2}{f_V}V_\mu A^\mu \ ,
\end{equation}
where the coupling constant $e/f_V$ can be determined by the vector meson leptonic decay $V\to
e^{+}e^{-}$,
\begin{equation}\label{vmd}
\frac{e}{f_V} = \left[\frac{3\Gamma_{V\to e^+ e^-}}{2\alpha_e
|p_e|}\right]^{1/2} \ ,
\end{equation}
where $|p_e|$ is the electron three-vector-moment in the vector meson rest
frame, and $\alpha_e = 1/137$ is the EM fine-structure constant. For the $\gamma-J/\psi$ coupling, we have  $e/f_{J/\psi}= 0.027$.

In the framework of VMD, the EM coupling $h_1$ in Eq.~(\ref{photocoupling}) can be related to the strong coupling in Eq.~(\ref{strong-coupling}):
\begin{equation}
e h_1= -\frac{e M_{J/\psi}^2}{f_{J/\psi}} \frac{i g_1}{k^2-M_{J/\psi}^2} = i \frac{e}{f_{J/\psi}} g_1 \ ,
\end{equation}
where the real photon condition $k^2=0$ is applied. The strong coupling constant $g_1$ can then be estimated by the $P_c^+\to J/\psi p$ data. If we assume $J/\psi p$ channels accounts for the total widths of $P_c^+(4380)$ and $P_c^+(4450)$, i.e.  $205\pm 18\pm 86$ MeV and $39\pm 5\pm 19$ MeV. We extract the coupling $g_1$ for different quantum numbers as
\begin{eqnarray}
g_{\frac 32^+}=1.07,\quad g_{\frac 32^-}=1.40,\quad g_{\frac 52^+}=2.56,\quad g_{\frac 52^-}=5.58 \ ,
\label{eq-coupling}
\end{eqnarray}
which will give the upper limit of their production cross sections in $J/\psi$ photoproduction. 

For states which are off-shell, we introduce a form factor as usually applied in hadronic models:
\begin{equation}
{\cal F}(p^2)=\frac{\Lambda^4}{\Lambda^4+(p^2-M_{P_c}^2)^2} \ ,
\end{equation}
where $p$ is the four-vector momentum carried by the $P_c$. Here, we use the value of $\Lambda=0.5 \mathrm{GeV}$ which is more suitable for heavier meson productions~\cite{Kim:2011rm}. One notices that the cross section from these two $P_c$ states will be even larger if $\Lambda$ increases.

It should be mentioned that in this scheme, we do not consider the nucleon Born terms or $t$-channel meson exchanges. Various studies have shown that these contributions are negligible in the energy region beyond the conventional $N^*$ region. This actually provide advantages for photoproduction reaction to directly produce the hidden-charm pentaquark in the $s$ channel which can be distinguished from the $t$-channel processes. In this formalism most of the parameters have been well constrained. The only parameter $g_1$ will have the upper limit given by the experiment as given in Eq.~(\ref{eq-coupling}).

\section{Numerical results}

\begin{figure}
\centering
\includegraphics[width=0.9\textwidth]{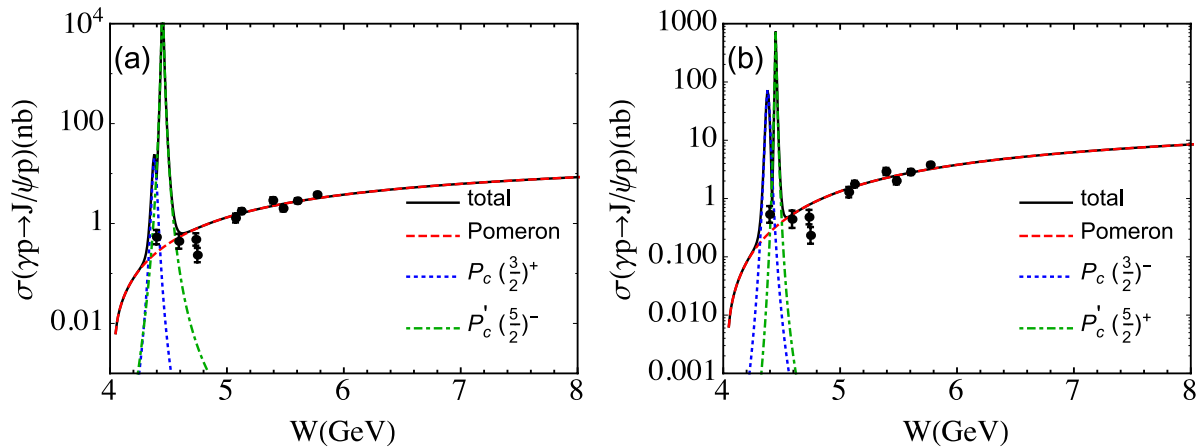}
\caption{The total cross section of $\gamma p\to J/\psi p$ in terms of the c.m. energy. The red dashed, blue dotted, green dot-dashed and black solid curves are the contributions from Pomeron, $P_c$ with spin $3/2$, $P_c$ with spin $5/2 $ and the coherent sum of all. (a) shows $(3/2^+, \ 5/2^-)$ combination. (b) shows $(3/2^-, \ 5/2^+)$ combination. The coupling constants (Eq.~(\ref{eq-coupling})) between $J/\psi p$ and the two $P_c$ states are extracted by assuming $J/\psi p$ saturates all their total widths. The experimental data are from Refs.~\cite{Camerini:1975cy,SLAC:1976,Gittelman:1975ix}}
\label{fig-4}
\end{figure}

\begin{figure}
\centering
\includegraphics[width=0.9\textwidth]{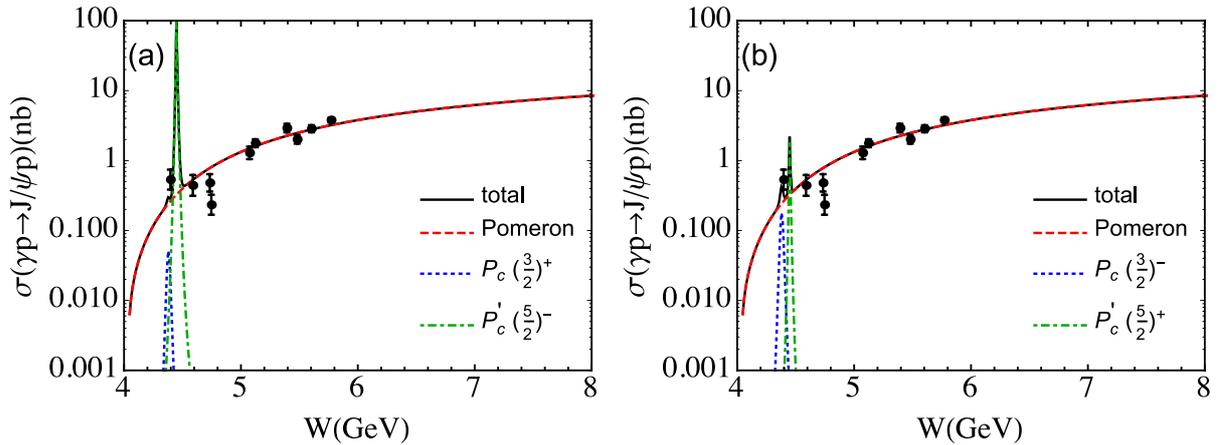}
\caption{The caption is the same as that of Fig.~\ref{fig-4} except that the couplings are extracted by assuming $J/\psi p$ channel only accounts for $5\%$ of their total widths.}
\label{fig-5}
\end{figure}

Proceeding to the numerical results, we first show  the total cross section from threshold to 8 GeV of the c.m. energy. The Pomeron exchange model can describe well the experimental data at high energies. The results are shown by the dashed line in Figs.~\ref{fig-4} and ~\ref{fig-5}. Then we argue that the extrapolation of the Pomeron exchange to the threshold region can still describe the behavior of the diffractive process in the $J/\psi$ photoproduction since the $t$-channel momentum transfer is still large at the forward angle. Although uncertainties of the Pomeron exchange model should exist near threshold, we emphasize that they should not change the smooth behavior of the total cross section. It means that interfering structures near threshold should indicate the contributions from non-diffractive processes such as the $s$ channel. 

Near threshold, the $s$-channel contributions from the pentaquark states can indeed bring obvious corrections to the total cross section as shown by the dotted and dot-dashed curves. In particular, the contribution from the narrow $P_c^+(4450)$ is significant. Notice that the data for the total cross sections near threshold still have relative large errors. The couplings between $J/\psi p$ and the two $P_c$ states in Fig.~\ref{fig-4} and Fig.~\ref{fig-5} are extracted by assuming it accounts for their total widths and $5\%$, respectively. The combined total cross section (solid line) in Fig.~\ref{fig-5} appears to be consistent to some extent with the data which means $J/\psi p$ channel is not the dominant decay channel for these two $P_c$ states. As shown in Fig.~\ref{fig-5} (b), the favored combination $(3/2^-, \  5/2^+)$ is also favored from the present photoproduction data.  Meanwhile,  a refined measurement of the cross section lineshape near the threshold region is required. 

The calculation results are quite informative. on the one hand, we find that if the $J/\psi p$ decay mode saturates the total widths of these two pentaquark candidates, their cross sections in $J/\psi$ photoproduction will significantly overestimate the experimental data. On the other hand, we find that a fraction of about $5\%$ for the $J/\psi p$ decay mode can be accommodated by the present photoproduction data. It means the these two pentaquark candidates, if exist, should not be a simple $J/\psi p$ system. They should have more stronger couplings to other channels, such as $\Sigma_c^* \bar{D}^{(*)}$, $\Lambda_c^* \bar{D}^{(*)}$, $\chi_{c1} N$, or $J/\psi N$ plus multi-pions.

\begin{figure}
\centering
\includegraphics[width=0.9\textwidth]{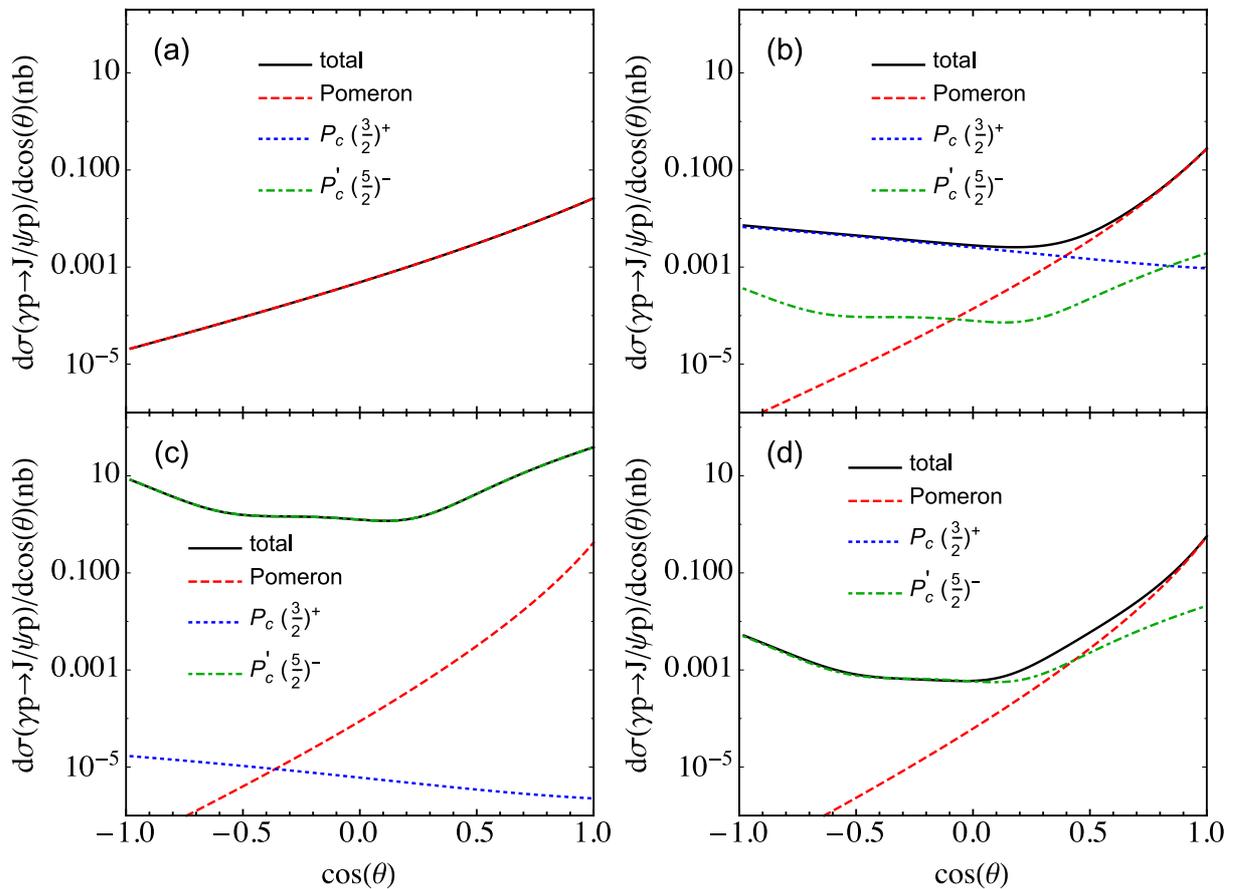}
\caption{Differential cross section of $\gamma p\to J/\psi p$ in terms of $\cos(\theta)$ with $\theta$ the relative angle between photon and $J/\psi$ in the c.m. frame.  The red dashed, blue dotted, green dot-dashed and black solid curves are the contributions from Pomeron, $P_c$ with spin $3/2$, $P_c$ with spin $5/2 $ and the coherent sum of all with the combination $(3/2^+, \ 5/2^-)$. (a), (b), (c) and (d) are at the center energy $W=4.15$ GeV, 4.38 GeV, 4.45 GeV and 4.5 GeV respectively.}
\label{fig-6}
\end{figure}

\begin{figure}
\centering
\includegraphics[width=0.9\textwidth]{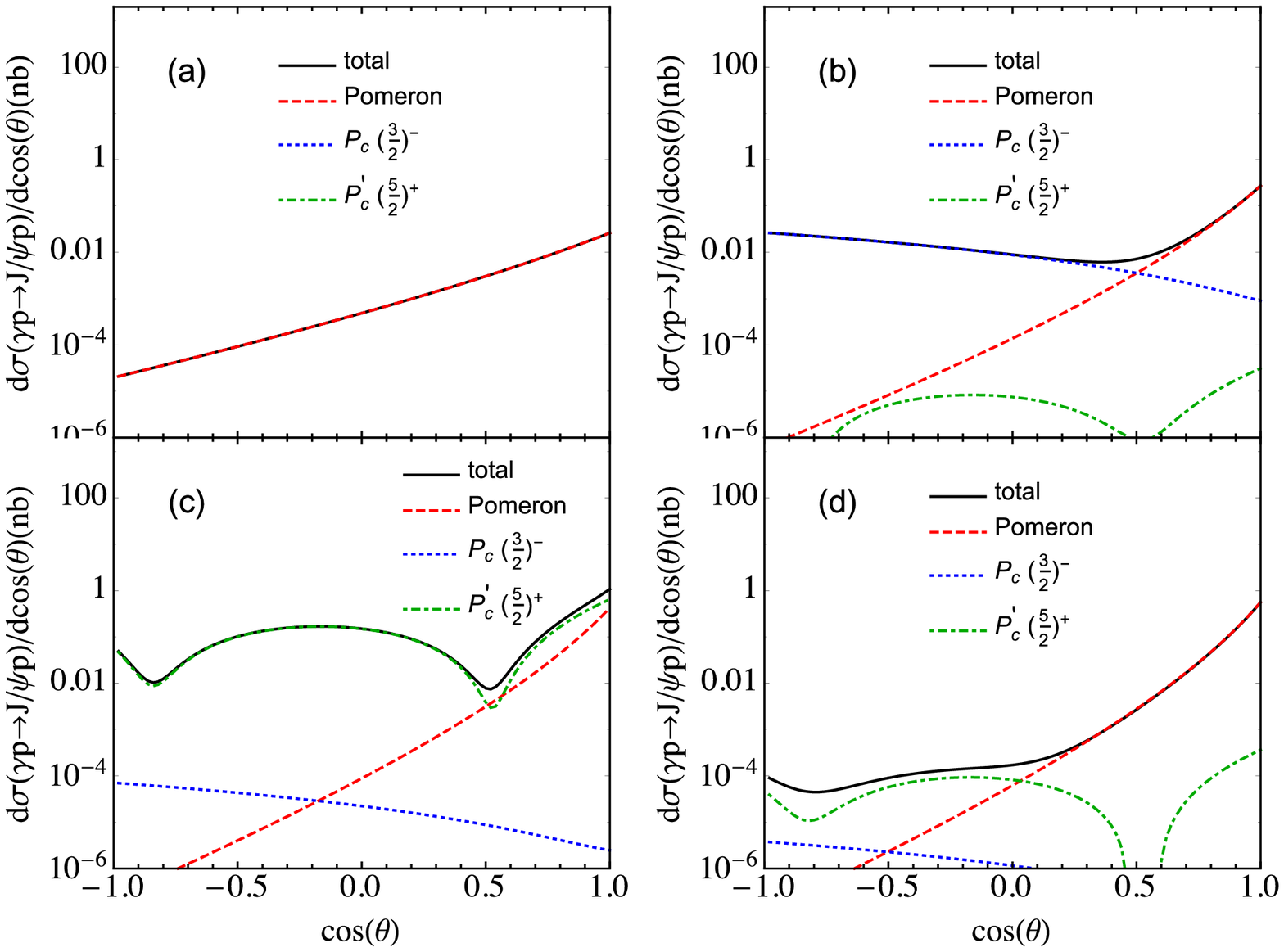}
\caption{Differential cross section of $\gamma p\to J/\psi p$ similar to Fig.~\ref{fig-6} except that quantum numbers of these two pentaquark states are $(3/2^-, \ 5/2^+)$. }
\label{fig-7}
\end{figure}

The angular distributions corresponding to Fig.~\ref{fig-5} at four energies, i.e. $W=4.15$ GeV, 4.38 GeV, 4.45 GeV and 4.5 GeV, are shown in Figs.~\ref{fig-6} and ~\ref{fig-7}, respectively. From Figs.~\ref{fig-6} (a), \ref{fig-6} (d), \ref{fig-7} (a), \ref{fig-7} (d), we can see that the $s$-channel contributions are restricted to a narrow kinematic region. Away from the $P_c$ states region it is still the Pomeron exchange plays a key role.  The differential cross section at 4.38 GeV and 4.45 GeV are dominanted by $P_c^+(4380)$ and $P_c^+(4450)$ as expected. The contributions from the two $P_c$ states make the differential cross section strongly deviated from the diffractive behavior as shown by Figs.~\ref{fig-6} (b), \ref{fig-6} (c), \ref{fig-7} (b), and \ref{fig-7} (c). These distributions can be measured by experiment at corresponding energies.

In Brief, either a detail scan of the total cross section in the low energy region or measurement of the angular distributions around 4.38 GeV and 4.45 GeV in $J/\psi$ photoproduction will help us pin down the nature of the two $P_c$ states. It should be pointed out, if any other possible states which can also couple to $J/\psi p$ strongly, they will also cause non-diffractive effects at off-forward angles which can be measured directly.

\section{Summary}

We propose to search for the newly observed hidden-charm pentaquark candidates $P_c^+(4380)$ and $P_c^+(4450)$ in $J/\psi$ photoproduction. If they are genuine states, their production in photoproduction should be a natural expectation. Since the cross section of the $J/\psi$ photoproduction is dominated by the diffractive process which is described by the $t$-channel Pomeron exchange, it should be relatively easy to isolate the direct production of these two states in the $s$-channel at off-forward angles in the differential cross section measurement. We find that the total cross sections near threshold appear to have a non-trivial lineshape due to the excitations of the hidden-charm pentaquark states. A detailed scan over the energy region from threshold to 5 GeV should be able to tell whether  $P_c^+(4380)$ and $P_c^+(4450)$ are genuine states or kinematic effects caused by the anomalous triangle singularity~\cite{Guo:2015umn,Liu:2015fea}.

\section*{Acknowledgement}

This work is supported, in part, by the  Sino-German CRC 110 ``Symmetries and
the Emergence of Structure in QCD" (NSFC Grant No. 11261130311), the National Natural Science Foundation of China (Grant Nos. 11035006 and 11425525), and the Japan Society for the Promotion of Science under Contract No. P14324 and the JSPS KAKENHI (Grant No. 25247036).

\end{document}